\documentclass[final,authoryear,times,1p]{elsarticle}

\usepackage{graphicx}
\usepackage{natbib}
\usepackage{multirow}
\usepackage[small]{caption}
\usepackage{color, colortbl}
\usepackage{subfig}
\usepackage{setspace}
\usepackage[titletoc,toc]{appendix} 
\usepackage{name ref} 
\usepackage{float}
\usepackage{longtable}
\usepackage{array} 
\usepackage{amsmath}
\usepackage{amssymb}
\usepackage{epsfig}
\usepackage{gensymb}
\usepackage{textcomp} 
\usepackage{siunitx}
\usepackage[euler]{textgreek}
\usepackage[usenames,dvipsnames,svgnames,table]{xcolor}
\usepackage{rotating}
\definecolor{Gray}{gray}{0.85}

\usepackage{rotating}
\usepackage{amssymb}
\usepackage[table]{xcolor}

\journal{Progress in Nuclear Energy}

\begin{document}
\newcommand{\keff}{k$_{\rm{eff}}$}
\begin{frontmatter}



\title{The Core Design of a Small Modular Pressurised Water Reactor for Commercial Marine Propulsion}


\author[man1,man2]{Aiden Peakman} \corref{cor1}\ead{a.peakman@liverpool.ac.uk}
\author[man1,man3]{Hywel Owen}
\author[man1]{Tim Abram}

\address[man1]{The University of Manchester, Manchester M13 9PL, UK}
\address[man2]{The University of Liverpool, Liverpool, Merseyside, L69 3GH, UK}
\address[man3]{The Cockcroft Institute, Daresbury WA4 4AD, UK}

\begin{abstract}

If international agreements regarding the need to significantly reduce greenhouse gas emissions are to be met then there is a high probability that the shipping industry will have to dramatically reduce its greenhouse gas emissions. For emission reductions from ships greater than around 40\% then alternatives to fossil fuels - such as nuclear energy - will very likely be required. A Small Modular Pressurised Water Reactor design has been developed specifically to meet the requirements of a large container ship with a power requirement of 110~MWe. Container ships have a number of requirements - including a small crew size and reduced outages associated with refuelling - that result in a greater focus on design simplifications, including the elimination of the chemical reactivity control system during power operation and a long core life. 

We have developed a novel, soluble-boron free, low power density core that does not require refuelling for 15 years. The neutronic and fuel performance behaviour of this system has been studied with conventional UO$_2$ fuel. The size of the pressure vessel has been limited to 3.5 metres in diameter. Furthermore, to ensure the survivability of the cladding material, the coolant outlet temperature has been reduced to 285$^{\circ}$C from 320$^{\circ}$C as in conventional GWe-class PWRs, with a resulting reduction in thermal efficiency to 25\%. The UO$_2$ core design was able to satisfactorily meet the majority of requirements placed upon the system assuming that fuel rod burnups can be limited to 100 GWd/tHM. The core developed here represents the first workable design of a commercial marine reactor using conventional fuel, which makes realistic the idea of using nuclear reactors for shipping.

\end{abstract}


\end{frontmatter}



\section{Introduction}

This study focuses on designing a reactor capable of powering a large container ship since container ships dominate greenhouse gas (GHG) emissions from the shipping industry (\cite{Vergara2012}). As a large proportion of a ship's operating cost is devoted to fuel costs there has already been a considerable drive to maximise fuel efficiency of conventional large diesel engines. Therefore beyond limiting ship speeds further, greater reductions in their GHG emissions are likely to be limited without changes to fuel types (\cite{Vergara2012}).

Whilst nuclear-powered ships have successfully operated at sea for a number of decades, these have been primarily naval systems (or derivatives of naval systems, such as icebreakers) and a few demonstration projects using reactors with low power outputs, typically ~50 MWth or less (\cite{Spyrou2006, WNA_Ships, Bukharin2006}). The operational requirements of large civilian vessels (for example high capacity factors and economic competitiveness) mean the naval and past demonstration reactor systems are ill-suited for use in the current fleet of commercial container ships. For instance, the fuels that have historically been utilised in naval systems are so-called ceramic-metallic (cermet) fuels and metallic dispersion fuels (\cite{Frost1982}). These fuel types have considerable drawbacks for use in a commercial setting primarily because of their relative expense compared to conventional fuel in Light Water Reactors (LWRs), and also because their low uranium densities necessitate the use of highly enriched uranium in order to achieve long core life (\cite{IAEA2003, Reeve1975}). The predominant reason for their historic use has been because of their excellent fuel performance characteristics (\cite{Frost1982}); for example in the case of ceramic particles dispersed in a metallic matrix, the fuel form has an effectively higher thermal conductivity and therefore operates at lower temperatures. This is beneficial as many fuel properties degrade at higher temperatures. Moreover, large temperature gradients across the fuel, with accompanying differences in thermal expansion behaviour across the fuel form can result in excessive stresses applied to the cladding material and ultimately clad failure. Nevertheless, conventional Pressurised Water Reactor (PWR) fuel in commercial reactors (pellets in a cylindrical Zircaloy tube) are relatively inexpensive and have demonstrated - through thousands of reactor years of operation, low failure rates (\cite{Uffelen2012}). Hence the preference in this study to employ conventional PWR fuel.

In recent years, so-called Small Modular Reactors (SMRs), with power outputs less than 300 MWe, have gained attention (\cite{OECD-NEA2011}). The primary aim of SMRs is to reduce the size of the reactor in order to reduce the capital cost to the extent where it becomes favourable for certain sectors to consider investing in nuclear power (many sectors cannot afford nor require $>$1~GWe plants as is the case in the shipping sector) and also to benefit from economies of mass production rather than economies of scale (\cite{OECD-NEA2011}). The relative trade-off between economies of scale vs economies of mass production is highly uncertain as usually there are a number of essentially fixed costs in building a nuclear power plant, for instance licensing the reactor and securing the facility; therefore, in a conventional nuclear power plant, specifying a larger power reduces the relative expense of these costs. Furthermore, for many reactor components it may be the case that their price does not scale directly with the power output  (\cite{OECD-NEA2011}). However, there are benefits that may offset some of these disadvantages; for example a smaller size may permit new reactor designs and manufacturing methods to be utilised (\cite{Ingersoll2009a}). As the vast majority of reactors that have been deployed at sea have been PWRs, we have decided here to focus on developing a PWR core specifically designed to power large container ships, with power demands of approximately ~100 MWe.

There is considerable uncertainty about the feasibility of being able to license a nuclear powered ship with a limited number of personnel on board (see Section \ref{sec:SBF operation} for more details). Hence it is prudent to try to identify systems that add to the complexity of the reactor system and whose removal may significantly simplify reactor operation, minimising demands upon personnel on board the ship. A particular focus in the present study was given to systems that directly influence neutronics and fuel performance. A key system identified was the system that delivers and maintains the correct concentration of Soluble Boron (SB).

\subsection{Non-Technical Barriers to Deployment}

There are a number of non-technical barriers to deploying nuclear-powered civilian vessels. Firstly, insurance presents a considerable problem. There is currently insufficient actuarial data to accurately determine the risks associated with operating nuclear-powered container ships (\cite{RAEng2013}). Whilst there are operating PWRs on board ships at sea, these do not operate under the same conditions as container ships, which frequently visit ports near high population centres and travel in busy ship lanes where the risk of collision is not negligible.  This results in a large degree of economic uncertainty with respect to ensuring sufficient finance is in place to cover the costs associated with any accident. In order to reduce the economic risks to the operator of the ship it may be necessary to insist on government (or governments) to underwrite the insurance. This is currently the case for conventional land-based nuclear power plants, where the operator's liability is typically limited to around $\pounds$1 billion, and beyond this limit the state acts as insurer of last resort (\cite{WNA_liability}).

Port restrictions are also likely to be in place as a number of countries are opposed to nuclear power and may not permit the vessels to enter their waters. However, the overall effect will likely be limitations with respect to  the route a nuclear-powered ship can take and result in the ship exhibiting a higher capacity factor, which has been factored into this study.

Disposing of nuclear waste will also very likely be more politically complex than for waste produced at land-based systems. This is because whilst the operator will likely have to cover the cost of disposal, it will also be necessary to find a country willing to host the nuclear waste. There is currently opposition within many countries regarding for disposal of nuclear waste from land-based systems even when those countries have benefited for decades from the low-carbon electricity their reactors have provided (\cite{Gill2014}). Hence, if the ship has served a number of countries in transporting goods, should the waste therefore be split between those countries or should the operator come to some sort of arrangement with their own country to deal with the waste? This is an open question that this study does not address. It is important to note however that technical solutions have been developed for safely disposing of nuclear waste (\cite{IAEA2006}), with the limited progress made internationally not due to lack of technical solutions but rather the political complexity of hosting facilities for nuclear waste sites.

Another potential issue will be the need and costs associated with well-trained staff able to operate a nuclear reactor on board a ship and issues regarding reactor behaviour in the event of collisions, foundering (the sinking of a ship due to taking on board water) and grounding of the ship. In the case of foundering, there has been some precedent, as a number of nuclear powered vessels have been sunk (\cite{Kursk_accident, IAEA2001a}). However, in all of the cases reported the reactor was successfully shut down and remained so (\cite{IAEA2001a}). In addition, it appears that given the high-integrity materials used for constructing nuclear reactors,  the nuclear fuel has not been openly exposed to seawater. All of these have resulted in relatively limited exposure to radionuclides (\cite{IAEA2001a}).  In addition, it is certainly the case that concerns are more likely to be raised if a ship sinks in relatively shallow water near a populated area, and hence retrieval of the core may be necessary. But even here there is experience relating to the retrieval of once-active nuclear cores from the seabed (\cite{IAEA2001a}).

\section{Core Parameters and Materials}

Table \ref{Tab:Core_params} summarises the core parameters with the following sections detailing the reasoning behind these parameters.

\begin{table}[h!]
\centering
\begin{tabular}{|c|c|}
\hline 
\bf{Parameter} & \bf{Value} \\ \hline
Thermal Output & 440 MWth \\ \hline
Power density & 40 kW/l \\ \hline
Thermal Efficiency & 25\% \\ \hline
Capacity Factor & 93.5\% \\ \hline
Fuel & UO$_2$\\ \hline
Lattice Type & Standard square 17 by17 assembly \\ \hline
Cladding & 2nd Generation Zr alloy (Zirlo)\\ \hline
RPV Outer Diameter & 3.5 m \\ \hline
Targeted Core Life & 15 years \\ \hline
Average Coolant Temperature & 270$^{\circ}$C \\ \hline
\end{tabular}
\caption{Summary of key parameters for the reactor being designed in this study.}
\label{Tab:Core_params}
\end{table}

\subsection{Core Power}

Surveying recent engine designs capable of powering large container ships resulted in the propulsion power requirement being set at 80~MW(e) (\cite{Wartsila_engines, MAN_engines}). Given that propulsion usually makes up around 75\% of ship power requirements, with the other 25\% (the so-called hotel load, which refers to power requirements not related to propulsion) being needed to power refrigeration, navigation and crew amenities, the total power requirement for the reactor studied in this work was set to 110~MWe. It has therefore been assumed that the reactor will produce electricity rather than powering the propeller directly simply because the hotel load is quite substantial. 


\subsection{Core Life and Capacity Factor}

Refuelling a reactor on board a ship is unlikely to be as rapid as the targeted refuelling period of less than 20~days large PWRs are able to achieve (\cite{AP1000_brochure}). This is because large PWRs are designed to be flooded and a crane is already inside the containment building that can readily transfer the individual assemblies from the reactor building to the spent fuel pond. It was decided therefore that even if refuelling does require a longer dry docking period than is normally the case, this would still be acceptable if refuelling the core is performed infrequently. In addition, it is likely that ports able to refuel nuclear-powered ships will be very limited in number and expensive to operate. Therefore the greater number of ships each port is able to service, the lower the economic penalty.  All of these points lead to a strong motivation to study the maximum achievable core life.

Modern ships typically have lives of around 30~years (\cite{CCC2011}). Over this 30~year period they may typically undergo routine maintenance during dry docking which currently takes place every 5~years; however, there is the possibility that this will be expanded to 7.5~years in the near future due to the economic cost of taking ships out of service (\cite{Lloyds2014}).  It was also assumed that it would not be possible to design a nuclear reactor using well established materials capable of achieving a core life of 30~years because various degradation mechanisms (such as corrosion) would render the reactor inoperable long before the 30~year core life was achieved. However, it was thought that a 15~year core life would be achievable. A 15~year core life is easily accommodated with dry docking intervals of 5 or 7.5~years, and also means that each ship would only need to undergo refuelling once during its life.

A typical container schedule was assumed to be 3~weeks at full power whilst travelling between ports and 2~days in port for cargo loading. Hence, targeting a 15~year core life and assuming that the power level of the ship when in port is reduced to around 25\% of total power (the hotel load) and when at sea the core is operating at full power, these assumptions result in a Capacity Factor (CF) of 93.5\%.

\subsection{Core Size}

SMRs typically have Reactor Pressure Vessels (RPVs) with diameters of less than 3.5 m (\cite{WNA_SMRs}). A standard 17$\times$17 assembly was chosen here for the assembly design; this avoids the uncertainties associated with implementing a new grid design (\cite{Song2007}). Standard 17$\times$17 assembly designs have dimensions of 21.42~cm by 21.42~cm and can accommodate fuel pins up to diameters of 1.06~cm (\cite{Todosow2004}). As the RPV's outer diameter is set to be less than 3.5~m, this places limitations on the permitted number of fuel assemblies within the core. With the 17$\times$17 assembly design, the maximum number of assemblies that could fit into such a RPV was 89. The outer diameter of this array of assemblies was 245.14 cm with the active length of the fuel rod set to 245.14~cm.

\subsection{Soluble-Boron-Free Operation}
\label{sec:SBF operation}

An important requirement for a commercial marine reactor is minimising capital, operation and maintenance costs in order for the system to successfully compete with other propulsion technologies.  By allowing the reactor to operate without soluble boron and rely more heavily on mechanical reactivity control it was envisaged these costs could be reduced, in addition to some improvements in safety performance.

In large GWe-class PWRs, soluble boron is used to: ensure satisfactory shutdown margins in all core states with fewer rod control cluster assemblies; allow for reactivity control throughout life without distorting power profiles; and reduce the residual poison penalty by relying less on solid burnable poisons (\cite{EPRI_SBF}). In large reactors the proportional cost attributed to soluble boron operation is relatively small (\cite{EPRI_SBF}). However, for smaller reactor systems any additional operation complexity that results in greater personnel requirements (i.e. crew size) is likely disadvantageous as personnel costs start to disproportionately affect total operation costs. For comparison, a large PWR will typically have around 500 to 800 staff working onsite (\cite{EDF_SxB}), whereas a modern container ship will have a crew size of around 13 people (\cite{ABS_maersk}).

The elimination of soluble boron during operation would allow for considerable simplifications of the Chemical Volume and Control System, in particular a closed loop, high-pressure coolant purification system could probably be employed (\cite{EPRI_SBF}). In addition, chemistry control would be significantly simplified and the nuclear sampling system would require less frequent samples to be taken. Furthermore, the likelihood of boric-acid induced corrosion occurring could be significantly reduced, thereby reducing the requirement for inspections and evaluation of boric-acid-induced corrosion (\cite{Fiorini1999}). There would also be significant benefits with respect to safety performance as the likelihood of boron dilution accidents would be greatly reduced as the core should very rarely operate with boron in the primary coolant (\cite{Fiorini1999}). 

We assume it would be difficult to entirely eliminate the requirement for soluble boron as in the event of any core damage resulting in fuel relocation it would be necessary to utilise an alternative means for ensuring satisfactory shutdown margins. Therefore an emergency boron system, such as is implemented in Boiling Water Reactors (BWRs), would be kept in reserve to act as a redundant and diverse means to ensure reactor shutdown.

\subsection{Operating Temperature}
\label{sec:operating_temp}
 
One of the important limiting phenomena in PWRs relates to corrosion of the fuel, that is the amount of oxidation and the extent to which liberated hydrogen during oxidation is absorbed by the cladding material resulting in embrittlement. Oxidation and hydrogen absorption are phenomena dependent on temperature and time, with correlations relating corrosion as a function of temperature and time constructed from empirical measurements (\cite{R.L.2012}). Commercial PWRs typically have core inlet and outlet temperatures around 290$^{\circ}$C and 320$^{\circ}$C respectively (\cite{WNI2012}), with typical fuel residence times in the core of around 5 years. It would be impractical to expect fuel with current advanced zirconium alloy cladding materials to last for 15 years in the core, operating at temperatures between 290$^{\circ}$C and 320$^{\circ}$C.  Therefore coolant temperature in the marine reactor must be lowered, to reduce the likelihood of excessive corrosion so as to ensure the survivability of the fuel during its 15 year residence time.

There is limited information in the open literature on the corrosion performance of advanced zirconium alloys for periods of time comparable to the targeted core life. Hence it was decided to use CANDU Pressure Tubes (PTs) as a proxy to estimate necessary coolant temperatures for the core in this study. PTs typically reside in CANDU reactors for decades, with the latest PT designs designed to be replaced every 30 years (\cite{R.L.2012}). Furthermore, given that CANDU systems operate without soluble boron it is envisaged that the chemistry (in particular the pH) will be broadly similar to the marine core designed in this study. However, PTs can only be used as a crude proxy for predicting the performance of advanced zirconium alloys in PWR conditions since:

\begin{itemize}
\item PTs utilise a different zirconium alloy to those in commercial PWRs;
\item temperature gradients are lower in PTs; and
\item neutron and gamma fluence is much lower than in PWR cladding material.
\end{itemize}

The information in \cite{B_Cox} implies zirconium cladding materials can survive for around 15 years if core inlet and outlet temperatures are limited to 255$^{\circ}$C and 285$^{\circ}$C, with peak clad temperatures limited to 310$^{\circ}$C. The low coolant outlet temperature compared with conventional PWRs will reduce the achievable thermal efficiency and it has been assumed that the thermal efficiency will be around 25\%. Given that the core must produce 110~MWe and has a CF of 93.5\% this results in an average thermal power of 411.4~MWth.

\subsection{Selected Burnable Poison}
\label{Selected_BP}

The focus of this study is employing, as far as is reasonably possible, conventional light water reactor technology in order to achieve a 15 year core life. Given that the system will operate without soluble boron it is necessary to utilise a higher poison loading than is typical in conventional PWRs. To date industry experience has focused on three Burnable Poisons (BPs):  Gd$_2$O$_3$, Er$_2$O$_3$ and ZrB$_2$.

ZrB$_2$ was ruled out for the reactor investigated here since there was concern that the high concentrations of ZrB$_2$ necessary would create issues regarding rod internal pressure from helium production (\cite{Hesketh2012}); also, the thickness of ZrB$_2$ necessary to suppress reactivity would be far beyond existing experience raising concerns about the mechanical behaviour of the layer. Er$_2$O$_3$ exhibits favourable neutronic properties for long-life cores but current experimental data indicates poor behaviour relating to achievable fuel densities at high concentrations ($>$ 2~wt.\%) (\cite{Yamanaka2009}). Gd$_2$O$_3$ does not suffer from helium production during irradiation and it there is prior experience with manufacturing (U, Gd)O$_2$ fuel with high Gd$_2$O$_3$ up to around 15~wt.\% (\cite{Wada1973, IAEA1995}). Therefore,  Gd$_2$O$_3$ was employed in this marine reactor design.


\subsection{Control Rod Materials}
\label{sec:CR_materials}

The control rods employed in this study must be capable of operating for extended periods of time, deep into the core since they, along with solid burnable poisons, will be the predominant means to control reactivity. Most PWRs utilise Ag-In-Cd (macroscopic absorption cross-section of 9.9~cm$^{-1}$) control rods; however, a large proportion of the rod worth comes a single isotope ($^{113}$Cd) which accounts for only a small proportion of the available nuclides within the Ag-In-Cd alloy. Therefore, it is to be expected that this single isotope will burn out relatively quickly, greatly diminishing the rod worth (\cite{Gosset1993}). This is not a major issue in conventional PWRs since the control rods do not penetrate deep into the core during normal operation. 

An alternative material for control rod material is hafnium, which has a relatively low macroscopic absorption cross-section (4.8~cm$^{-1}$) relative to boron carbide (81~cm$^{-1}$), but it has superior properties under irradiation (low depletion of rod worth as a function of burnup, no gas release and limited swelling). Boron carbide has a much higher thermal absorption cross-section and in comparison to hafnium is relatively inexpensive. Hence, it was decided to employ a hybrid design whereby the lower portion that is consistently in the active region of the core is made of hafnium, with the other portion made of boron carbide to increase overall rod worth when control rods are fully inserted. This same strategy is successfully employed in BWRs (\cite{Horn2012}). The hybrid hafnium and boron carbide control rods employed in this design are in the standard 24 finger rod control cluster assembly.





\section{Comparison with other PWR systems}

There are a number of Small Modular PWRs (SM-PWRs) at various stages of development. Notable examples include: NuScale's SMR design, with a power output (factoring in the recently reported up-rate) of around 200~MWth/60~MWe and refuelling period of 2 years \citep{NUSCALE-2018}; the ACP100, with a power output of 310~MWth/100~MWe and refuelling period of 2 years \citep{Zhu2016}; and Westinghouse's SM-PWR concept, with an output of around 800~MWth/225~MWe \citep{Liao2016}. In common with the majority of SM-PWR designs, these SM-PWRs all nominally employ soluble boron in the coolant during normal operation. There are however also some concepts that intend to operate without soluble boron, for example the Flexblue SMR with an output of 550~MWth/160~MWe, and with some Flexblue core variants having a cycle length of 38 months \citep{FLEX-2015}. All of these systems have refuelling periods much less than the 15 years refuelling period we aim for. Furthermore, they do not specifically target the power demands of container ships. However, many of the techniques of modularisation these systems intend to employ, would directly benefit the system developed here, including advanced manufacturing techniques such as hot isostatic pressing and electron beam welding \citep{EPRI-2016}.

In contrast to large and small land-based PWRs, the core designed here will need to consider the unique characteristics of operating at sea. From a neutronic perspective, for a PWR system there are no direct impediments to operating at sea under normal operating conditions, i.e. the pitch and roll of the ship does not need to be taken into account in the calculations presented here. However, there are important characteristics such as shutdown and thermal-hydraulic characteristics that will require careful consideration, and would need to factored into the next stages of development for the marine core outlined here. In the case of shutdown, it is important that the control rods can be inserted and remain inserted independent of core orientation and thermal-hydraulics will need to ensure adequate cooling (including decay heat removal) under any orientation. Finally, in the event of grounding any systems that are nominally cooled by sea-water will necessitate alternative backup coolant options.

\section{Core design}
\subsection{Neutronics}

The core design employed Studsvik's CASMO-4/SIMULATE-3 neutronic suite (\cite{Arizona_Physics}) with the JEF 2.2 nuclear data library. CASMO-4 is a multi-group two-dimensional neutron transport theory code for modelling fuel assemblies, whereas SIMULATE-3 is a 3D nodal code that utilises diffusion theory. SIMULATE employs the output from CASMO-4 to model the entire reactor core and is used for determining the spatial and time dependence of the neutron flux throughout core life, where the slow variation of the global flux permits the use of the diffusion equation (\cite{Stammler}).

The core design process employed in this study consists of two key heuristic steps:

\begin{enumerate}
\item Optimisation with homogeneous fuel rods (no axial variation of enrichment or burnable poison) and no control rods present. Initially a loading pattern is developed with a relative power fraction (peak power to average power) highest in the centre of the core with a value below 1.5 at the beginning of life. Once this criteria is met, the optimisation proceeds to step 2.
\item Optimisation with heterogeneous fuel rods (axial variation of enrichment and poison concentration) and with control rods inserted into the core. A control rod loading pattern is developed where no Rod Control Cluster Assembly (RCCA) is permitted to be inserted greater than 50\% of the way into the core during normal operation. In addition, the control rods must meet a hot zero power shutdown criterion and a cold zero power (CZP) shutdown criterion. As will be shown, the CZP criterion is most difficult to meet and necessitated numerous iterations for the soluble boron free core designed here.
\end{enumerate}

\subsubsection{Optimisation with homogeneous fuel rods and no control rods present}

\begin{figure}[H]
  \centering
	\subfloat[Assembly with 76 burnable poison pins - this layout is utilised in later assembly designs.]{\label{Assembly_picture_76}\includegraphics[width=0.45\textwidth]{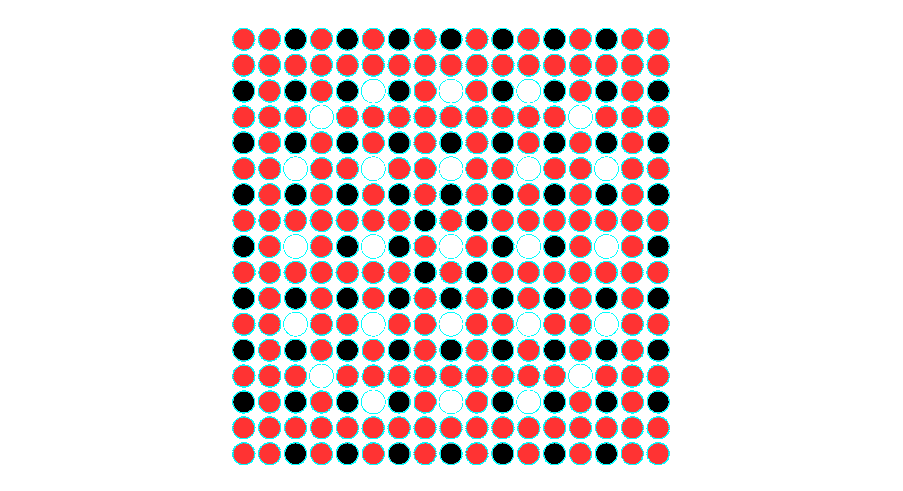}} 
  \hspace{0.1cm}               
  \subfloat[Assembly with 96 burnable poison pins.]{\label{Assembly_picture_96}\includegraphics[width=0.5\textwidth]{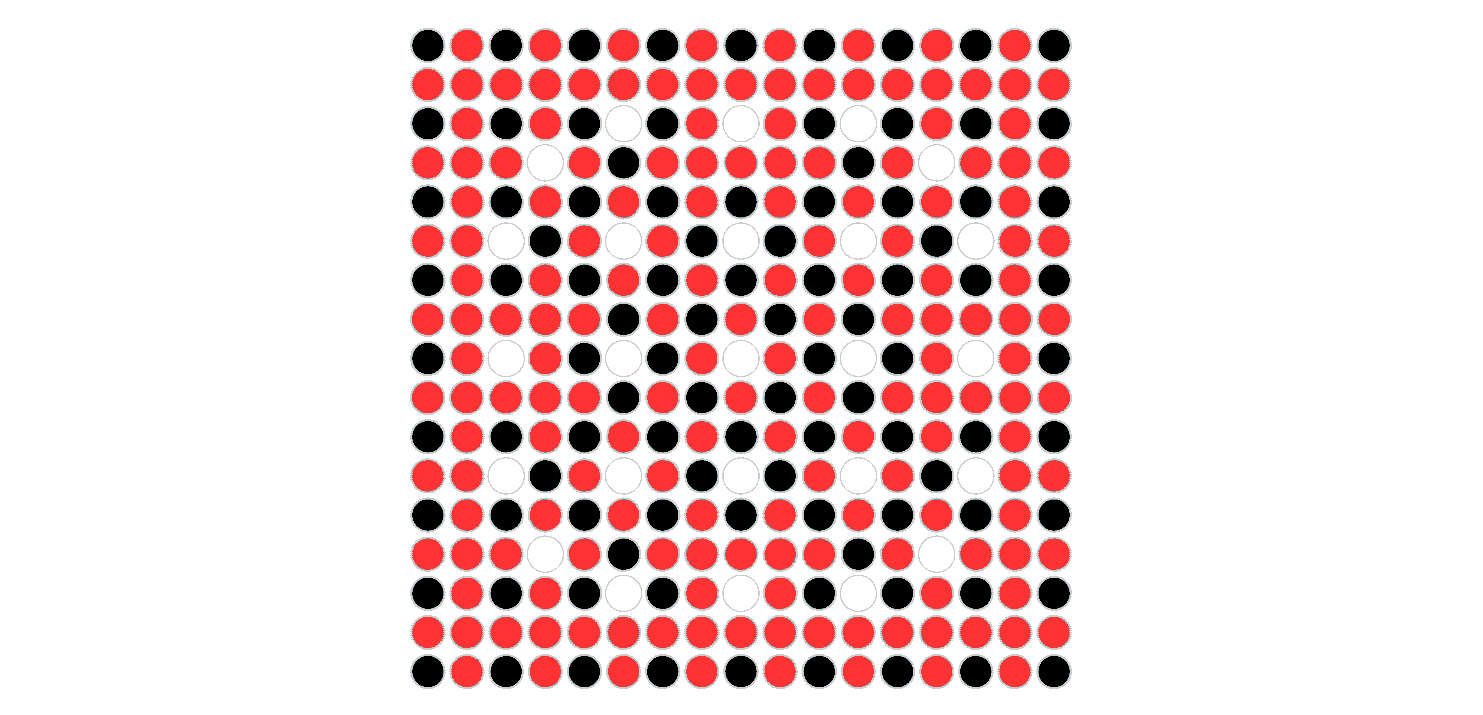}}
  \caption{Location of the Burnable Poison Pins (BPPs) (in black) which contain a mixture of UO$_2$ and Gd$_2$O$_3$, and the fuel pins (in red). The layout of the BPPs was based on the fact that the reactivity, and therefore power, tends to peak in fuel rods next to the water channels where increased neutron moderation is taking place. Hence, BPPs are located in close proximity to these channels in order to try and limit power peaking in the assemblies.}
  \label{Assemblies}
\end{figure}

\begin{figure}[H]
\centering
\includegraphics[width=0.45\textwidth]{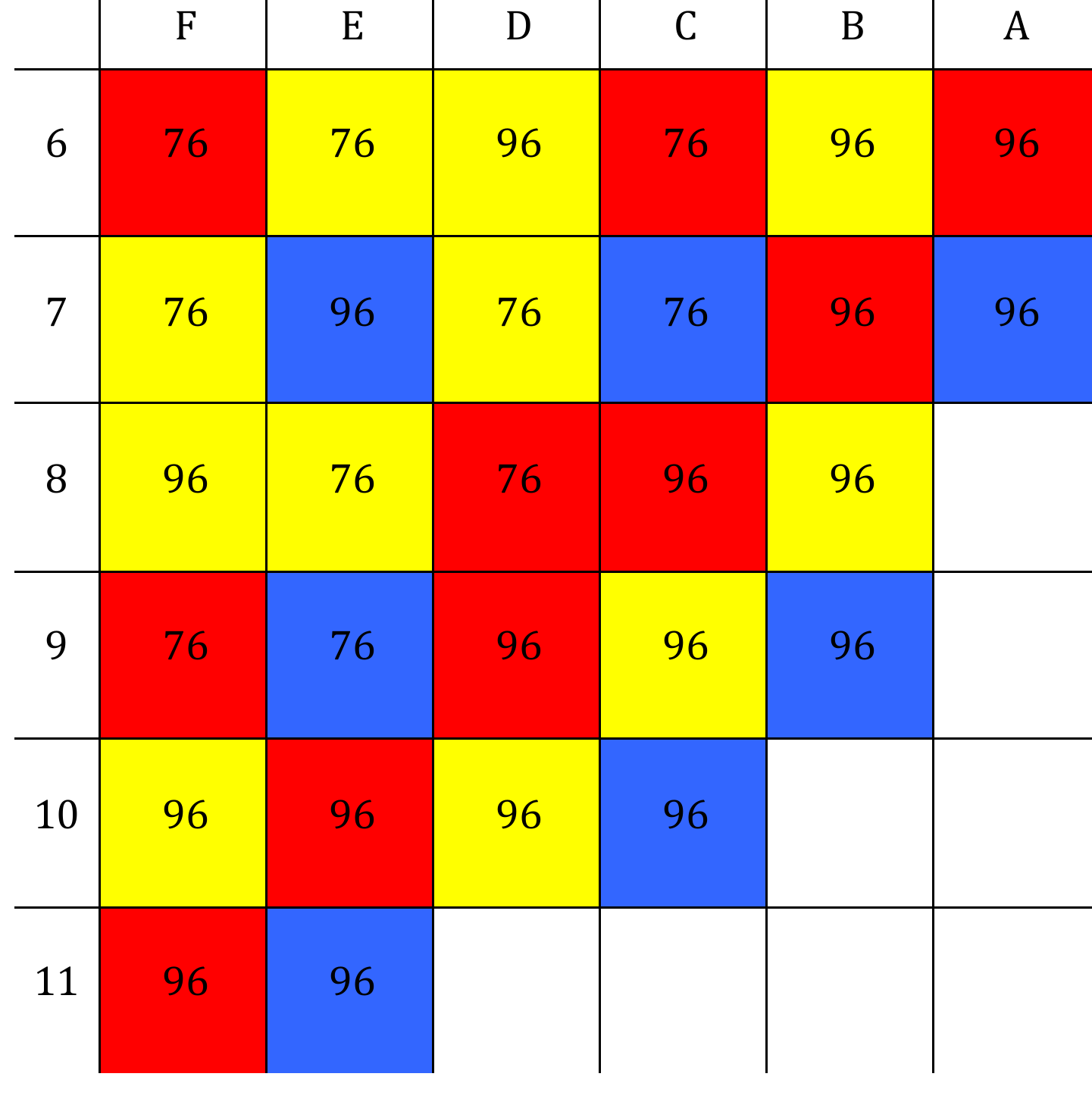}
\caption{A 1/4 core layout for the core designed in this study. The colours and numbers represent assemblies IDs with their contents shown in Table \ref{core_layout_breakdown}. }
\label{Core_layout}
\end{figure}

\begin{figure}[H]
\centering
\includegraphics[width=0.75\textwidth]{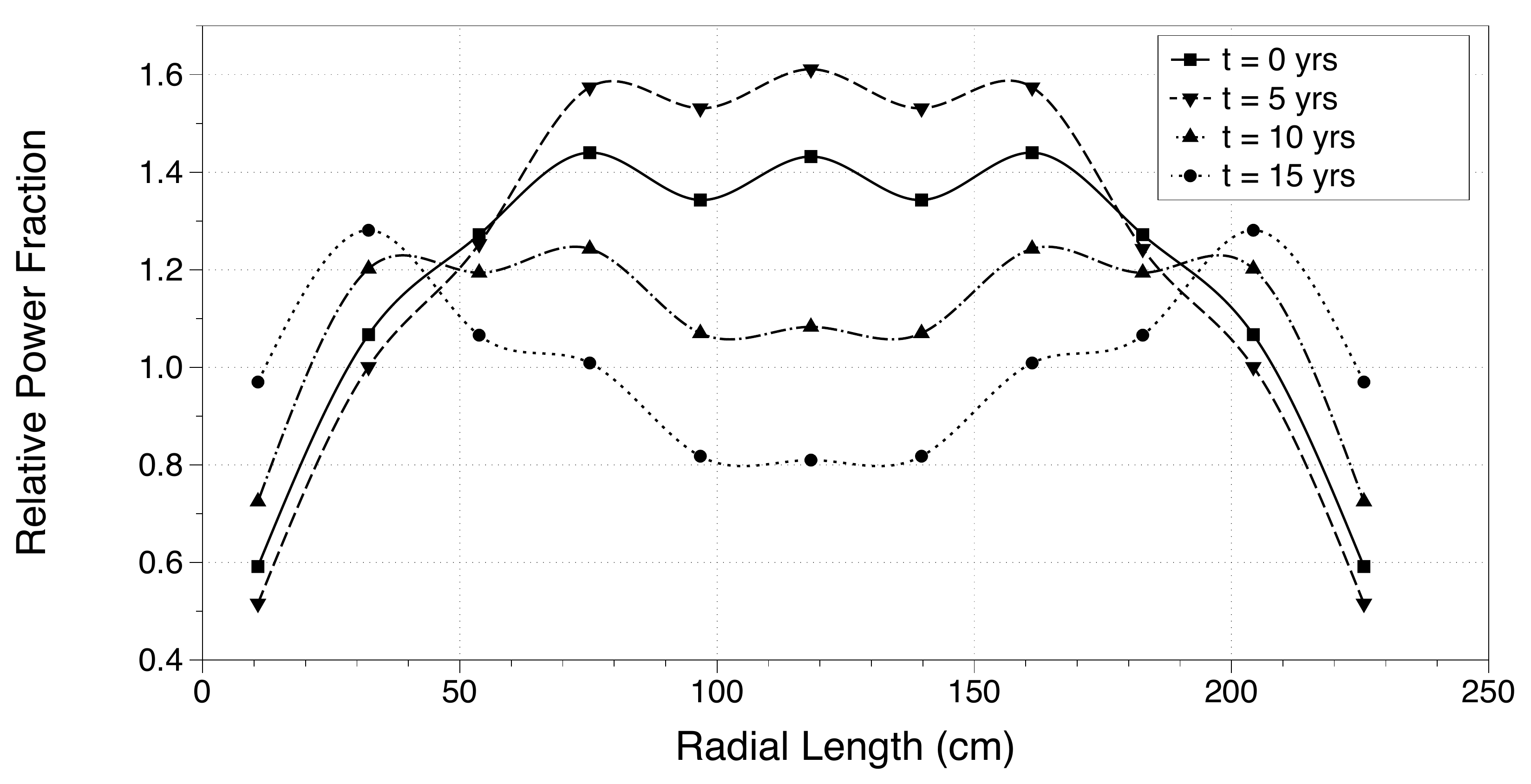}
\caption{Radial power profiles for core layout shown in Figure \ref{Core_layout} as a function of time, with a core diameter of 2.45 m. }
\label{2RPF_UO2_evolution}
\end{figure}

\begin{figure}[H]
\centering
\includegraphics[width=0.70\textwidth]{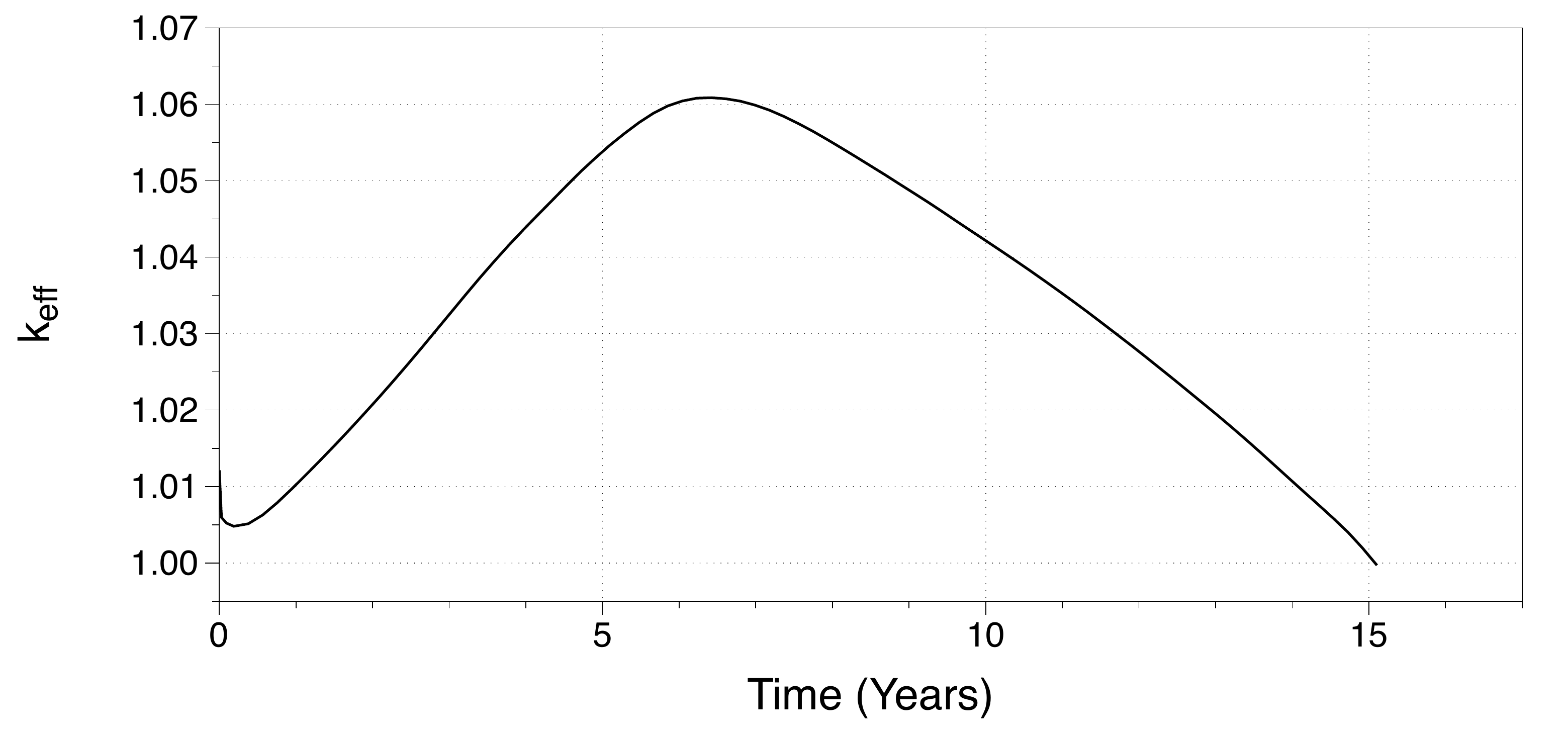}
\caption{k$_{\rm{eff}}$ evolution for core shown in Figure \ref{Core_layout} with homogeneous fuel rods and no control rods inserted.}
\label{LP3_8_iii_evolution}
\end{figure}

Due to the long core life and lack of soluble boron during normal operation, a large number of burnable poison pins (BPPs) were required in order to satisfactorily suppress reactivity over core life. Two distinct BPP layouts were chosen, with the ratio of these lattice types within the core adjusted to minimise excess reactivity and power peaking factors. Figure \ref{Assemblies} shows the two lattice types  and Figure \ref{Core_layout} shows the chosen layout for these lattices, based on manually optimising the position of the lattices to minimise power peaking. Table \ref{core_layout_breakdown} gives further details on the contents ($^{235}$U enrichment and Gd$_2$O$_3$ concentrations) of the lattice shown in Figure \ref{Core_layout}.  The limits in Table \ref{core_layout_breakdown} were chosen to be within the constraints outlined in Section \ref{Selected_BP} and obeying the 20 wt.\% limit on $^{235}$U content. Figures \ref{2RPF_UO2_evolution} and \ref{LP3_8_iii_evolution} shows the radial power profiles and reactivity variation as a function of time, respectively, for the core shown in Figure \ref{Core_layout}.

{\footnotesize
\tabcolsep=0.11cm
\begin{table}[h!]
\centering
\begin{tabular}{|c|c|c|}
\hline
\bf{Assembly ID} & \bf{Enrichment (wt.\%)} & \bf{Gd$_2$O$_3$ content (wt.\%)}\\ \hline
\cellcolor{red}96 & 16.5 & 11 \\ \hline
\cellcolor{yellow}96 & 15.5 & 12 \\ \hline
\cellcolor{blue!50}96 & 13.5 & 13 \\ \hline
\cellcolor{red}76 & 12.5 & 15 \\ \hline
\cellcolor{yellow}76 & 11.5 & 16 \\ \hline
\cellcolor{blue!50}76 & 10.5 & 17 \\ 
\hline
\end{tabular}
\caption{Breakdown of individual assembly contents. The assembly IDs correspond to those in Figure \ref{Core_layout}, with the number in the assembly ID describing the number of BPPs within that assembly. Note that pins within an assembly have identical enrichment but some pins will contain burnable poison  (at the concentration detailed above) and the remaining pins will consist of just UO$_2$.}
\label{core_layout_breakdown}
\end{table}%
}

\subsubsection{Optimisation with heterogeneous fuel rods and control rods present}

To compensate for the excess reactivity shown in Figure \ref{LP3_8_iii_evolution} a control rod arrangement needed to be chosen. Initially two key constraints were chosen that the control rods had to meet. These were:

\begin{itemize}
\item No RCCA must have an insertion depth $>$ 50\% during power operation - this heuristic rule was set as deep RCCA insertion during operation would likely result in large reactivity insertions in the event an RCCA is rapidly withdrawn during a transient;
\item The effective neutron multiplication factor (\keff) with highest worth rod stuck $<$ 0.98 at Hot Zero Power (HZP)
\end{itemize}

\begin{figure}[H]
\centering
\includegraphics[width=0.275\textwidth]{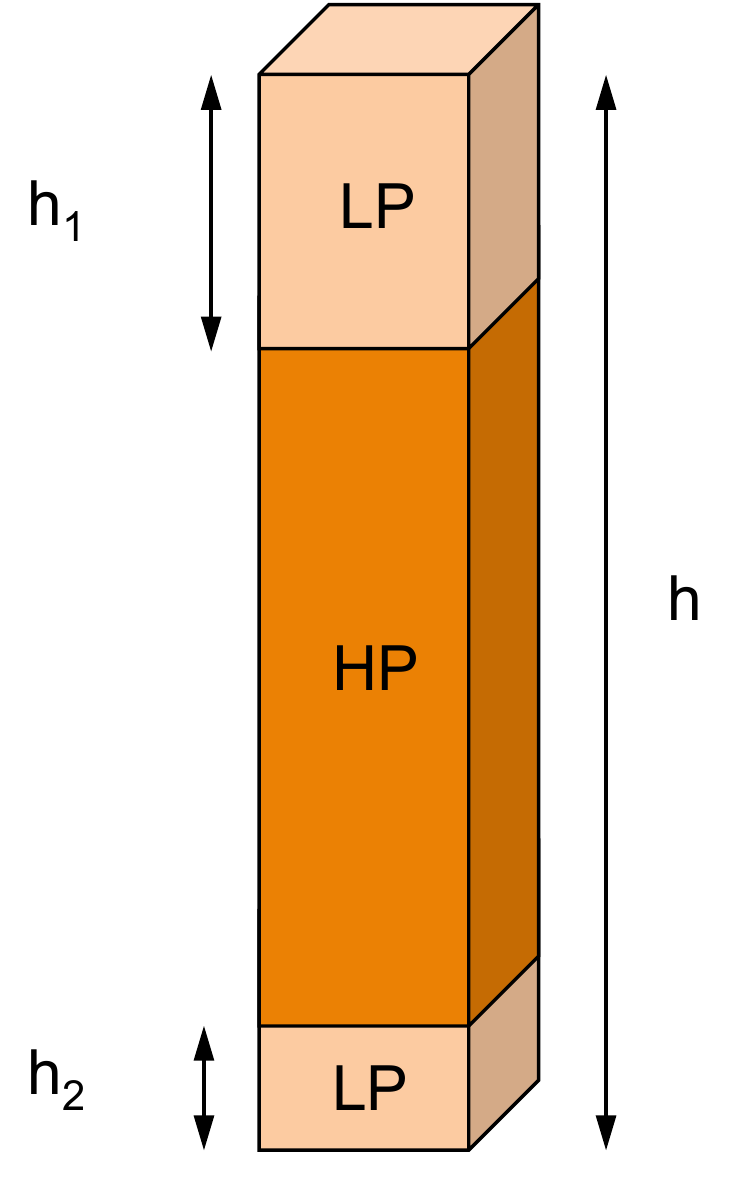}
\caption{Illustrating how the assemblies were split into 3 axial regions: a top Low Poison (LP) region, a middle High Poison (HP) region and a bottom LP region, to reduce power peaking when control rods were inserted. The colours indicate regions with the same enrichment and containing burnable poison pins with the same concentration of Gd$_2$O$_3$. The height of the top (h$_1$) and bottom (h$_2$) axial regions were determined as outlined in this section. }
\label{axial_diag}
\end{figure}

It was necessary to vary the poison and enrichment along the length of the assemblies to maximise rod worth and therefore minimise rod insertion. Starting from an axially homogenous design, the assemblies were divided into three regions as shown in Figure \ref{axial_diag}. The top and bottom region of each assembly have identical composition with lower poison concentration and/or higher $^{235}$U enrichment than the middle region. This will in general allow for an increase in rod worth and also flatter power profiles than an axially homogenous design.

The chosen axial enrichment/poison contribution is shown in Table \ref{Lattice_breakdown} and the values for h$_1$ and h$_2$ were 100~cm and 20~cm respectively.

{\footnotesize
\tabcolsep=0.11cm
\begin{table}[H]
\centering
\begin{tabular}{|c|c|c|c|c|}
\hline
\bf{Lattice} & \bf{Low Enrichment} &  \bf{High Enrichment} & \bf{High Poison} & \bf{Low Poison}\\
\bf{key} & \bf{region (wt.\%)} & \bf{region (wt.\%)} & \bf{content (wt.\%)} & \bf{content (wt.\%)}\\ \hline
\cellcolor{red}96 & 15.5 & 17.25 & 16 & 6.75 \\ \hline
\cellcolor{yellow}96 & 14.5 & 16.25 & 17 & 7.75 \\ \hline
\cellcolor{blue!50}96 & 12.5 & 14.25 & 18 & 8.75 \\ \hline
\cellcolor{red}76 & 11.5 & 13.25 & 18 & 10.75 \\ \hline
\cellcolor{yellow}76 & 10.5 & 12.25 & 19 & 11.75 \\ \hline
\cellcolor{blue!50}76 & 9.5 & 11.25 & 20 & 12.75 \\ 
\hline
\end{tabular}
\caption{Breakdown of individual lattice contents of the optimised core. The lattice keys correspond to those in Figure \ref{Core_layout}.}
\label{Lattice_breakdown}
\end{table}%
}

\begin{figure}[H]
  \centering
	\subfloat[Initial layout with 45 RCCAs.]{\label{RCCA_layout_1}\includegraphics[width=0.4\textwidth]{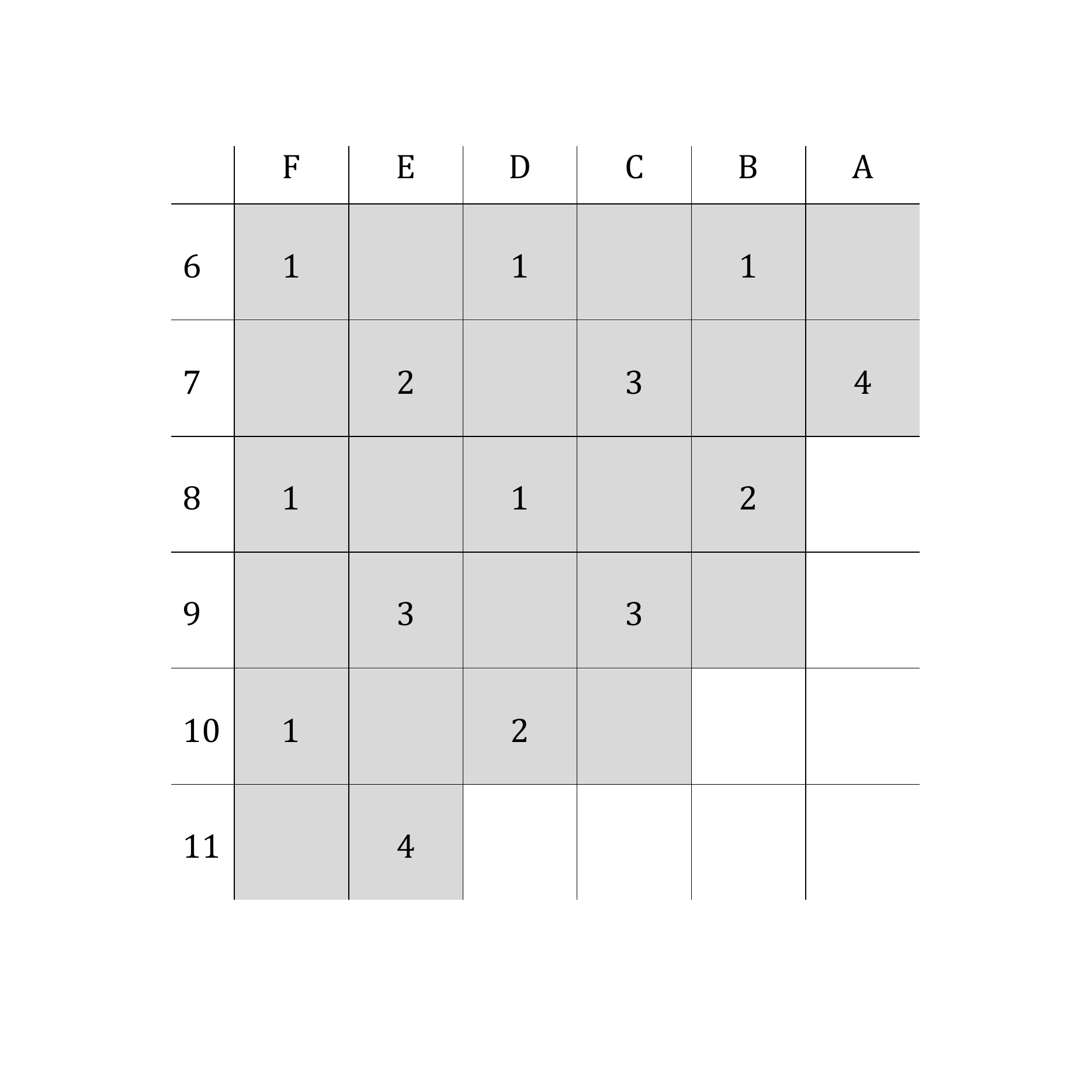}} 
  \hspace{0.4cm}                           
  	\subfloat[Second iteration of layout with 81 RCCAs.]{\label{RCCA_layout_2}\includegraphics[width=0.4\textwidth]{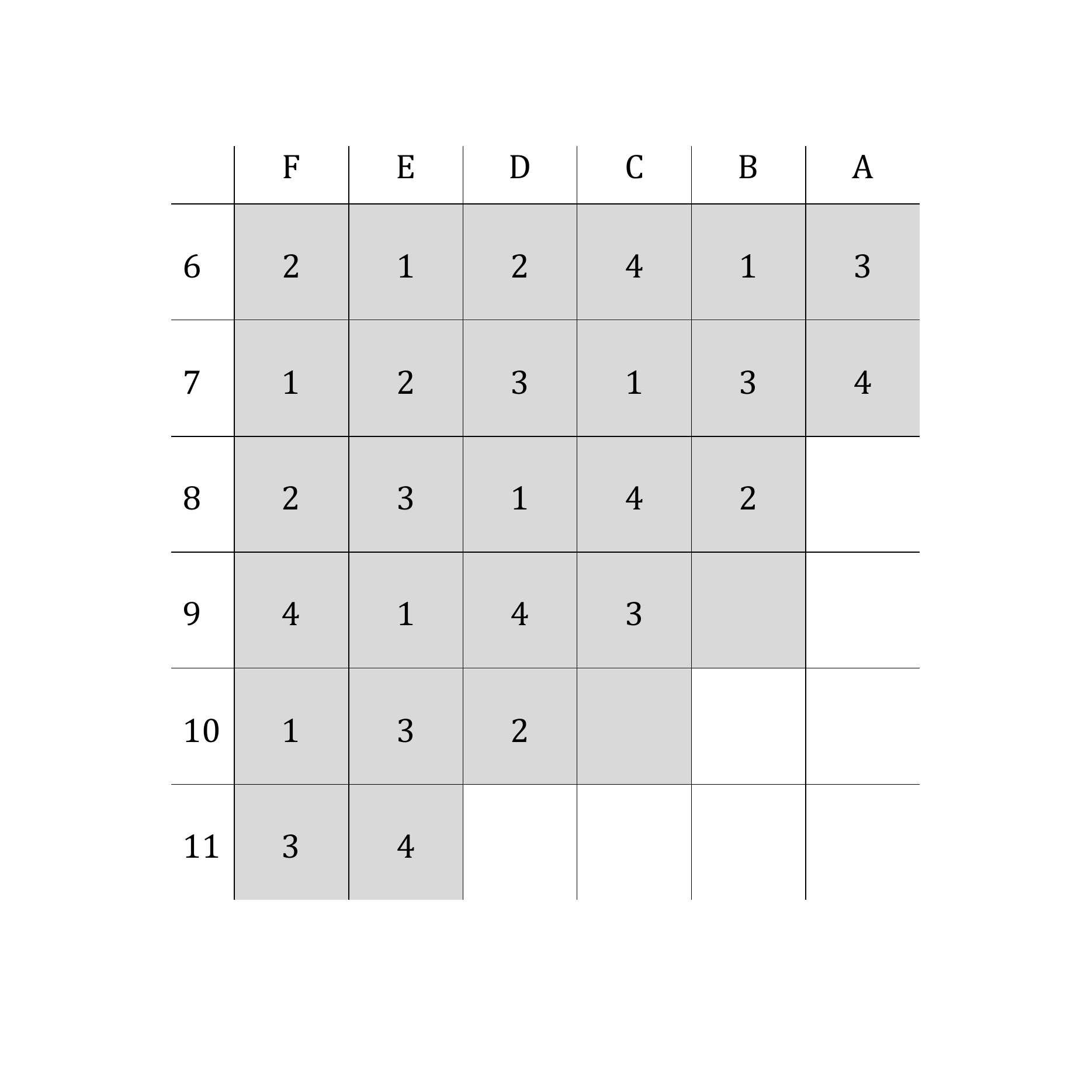}}
  \caption{The Rod Control Cluster Assemblies (RCCAs) groups (banks) for the first two configurations studied.}
  \label{RCCA_layouts}
\end{figure}

Initially a control rod loading pattern was selected (see Figure \ref{RCCA_layout_1}) and a matrix of poison contents, $^{235}$U enrichments and zone heights were chosen. A large number of cores were then modelled from these variables and the core that: 1) met the criteria relating to rod insertion limits during operation and the Hot Zero Power (HZP) shutdown criterion highlighted earlier: and 2) had the lowest through-life peaking factor, was selected as the most viable design. However, initial control rod loading patterns (with 45 RCCAs) were unsuccessful in meeting the needed shutdown margins and rod ejection criteria. Figure \ref{RCCA_layout_2} shows the optimised control rod loading pattern, with 81 RCCAs, that successfully met these requirements.

\begin{figure}[H]
\centering
\includegraphics[width=0.95\textwidth]{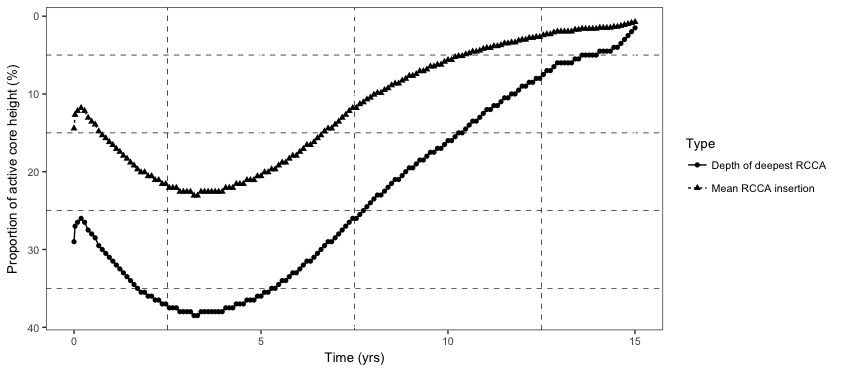}
\caption{Variation of RCCA position as a function of core life.}
\label{CR_details}
\end{figure}

Figure \ref{CR_details} details the variation of RCCA position as a function of core life for the axially heterogeneous core. Figure \ref{CR_details} shows that during power operation no control rods penetrate any more than around 40\% of the active core length. As discussed earlier, hafnium is well-suited for residing within the active region of the core for longer periods of time (it depletes slowly and does not exhibit unfavourable irradiation properties such as excessive swelling or helium production).  However, since hafnium is only relatively weakly absorbing, a hybrid RCCA design was employed with the remainder (60\%) of the RCCA consisting of B$_4$C for the reasons discussed earlier. The use of boron in the remainder of the rod will help meet the more stringent cold zero power (CZP) criterion of having a \keff \ with highest worth rod stuck $<$ 0.95 at Cold Zero Power (CZP).

The HZP shutdown criterion was satisfactorily met with the RCCA layout shown in Figure \ref{RCCA_layout_1}  (max \keff \ over core life under HZP conditions was 0.90). However with 81 RCCAs max \keff \ at CZP with highest worth rod stuck resulted in \keff  \ $>$0.95. Therefore, in the final iteration of the RCCA layout, a fifth bank, consisting of 8 RCCAs, was introduced (so that every fuel assembly had a dedicated RCCA) resulting in a maximum \keff \ of 0.92 at CZP with the highest worth rod stuck. This extra group of RCCAs would not need to penetrate the active region of the core during normal operation and would operate solely as a shutdown bank.

\subsubsection{Reactivity Coefficients and Xenon Transients}

An important characteristic of any viable core design is its stability to power changes. To ensure an inherently stable core design, the power coefficient of reactivity must be kept negative for all conceivable operating conditions. The two dominant effects in most reactors are the Doppler coefficient and Moderator Temperature Coefficient (MTC) (\cite{Duderstadt}). When licensing a core design the whole gamut of reactivity coefficients must be studied, but to simplify analysis only the MTC and Doppler coefficient were considered here. The Moderator Temperature Coefficients throughout life are in the range -5 to -76~pcm/$\,^{\circ}\mathrm{C}$. The Doppler coefficients were in the range -4 to -62~pcm/$\,^{\circ}\mathrm{C}$. These coefficients are consistent with those in large licensed PWRs that exhibit acceptable feedback behaviour (\cite{EDF2012}).

In conventional (GWe) LWRs, so-called xenon transients complicate startup, shutdown and power level changes. In the case of reactor shutdown, $^{135}$Xe will initially build up as the destruction mechanism of $^{135}$Xe to $^{136}$Xe is no longer available due to zero neutron flux, and the inventory of $^{135}$I will decay to $^{135}$Xe. The increase in $^{135}$Xe concentration implies that excess reactivity will be required to achieve re-criticality, as the inventory of $^{135}$Xe increases up until the point where the predominant mechanism determining $^{135}$Xe concentration is the decay of this isotope. Xenon transients are usually overcome using chemical reactivity control (i.e. adjustment in soluble poison concentration) and most reactors world-wide only undergo relatively modest changes in power or operate at a fixed power level, therefore xenon transients are only normally problematic in unplanned outages. However, for a marine reactor, there is no scope to operate the reactor at a fixed power level as the ship will enter port on a relatively frequent basis and the power level will need to be decreased. Therefore the behaviour of the core with power level changes was studied.

Table \ref{Xe_transients} shows the series of xenon transients performed on the core at various points in core life. Xenon clearly has minimal impact on core reactivity throughout life - for comparison conventional PWRs can experience negative reactivity insertions of up to approximately 1000~pcm. The reason behind the core's relative insensitivity to xenon transients (which is advantageous for a reactor expected to routinely undergo significant power changes) is due to the low thermal flux in the core, since the behaviour of xenon is heavily dependent on the steady-state flux level (\cite{Duderstadt}). The average thermal flux across the core within this reactor at end of life, calculated by SIMULATE, was found to be around 6$\times$10$^{12}$ neutrons s$^{-1}$cm$^{-2}$. For large PWRs the thermal flux is typically of the order of 10$^{14}$ neutrons s$^{-1}$cm$^{-2}$.

{\footnotesize
\tabcolsep=0.11cm
\begin{table}[h!]
\centering
\begin{tabular}{|c|c|c|}
\hline
\bf{Time transient} & \bf{Peak Xe} & \bf{Associated reactivity} \\
\bf{initiated} & \bf{time} & \bf{decrement} \\ \hline
10 d & 1.6~h & -6~pcm \\ \hline
4.90 y & 1.8~h & -11~pcm \\ \hline
10.0 y & 1.6~h & -10~pcm \\ \hline
15.0 y & 3~h & -38~pcm \\
\hline
\end{tabular}
\caption{Details the negative reactivity insertion associated with xenon transients taking place at various points in core life under Hot Zero Power conditions.}
\label{Xe_transients}
\end{table}%
}

\begin{figure}[H]
\centering
\includegraphics[width=0.85\textwidth]{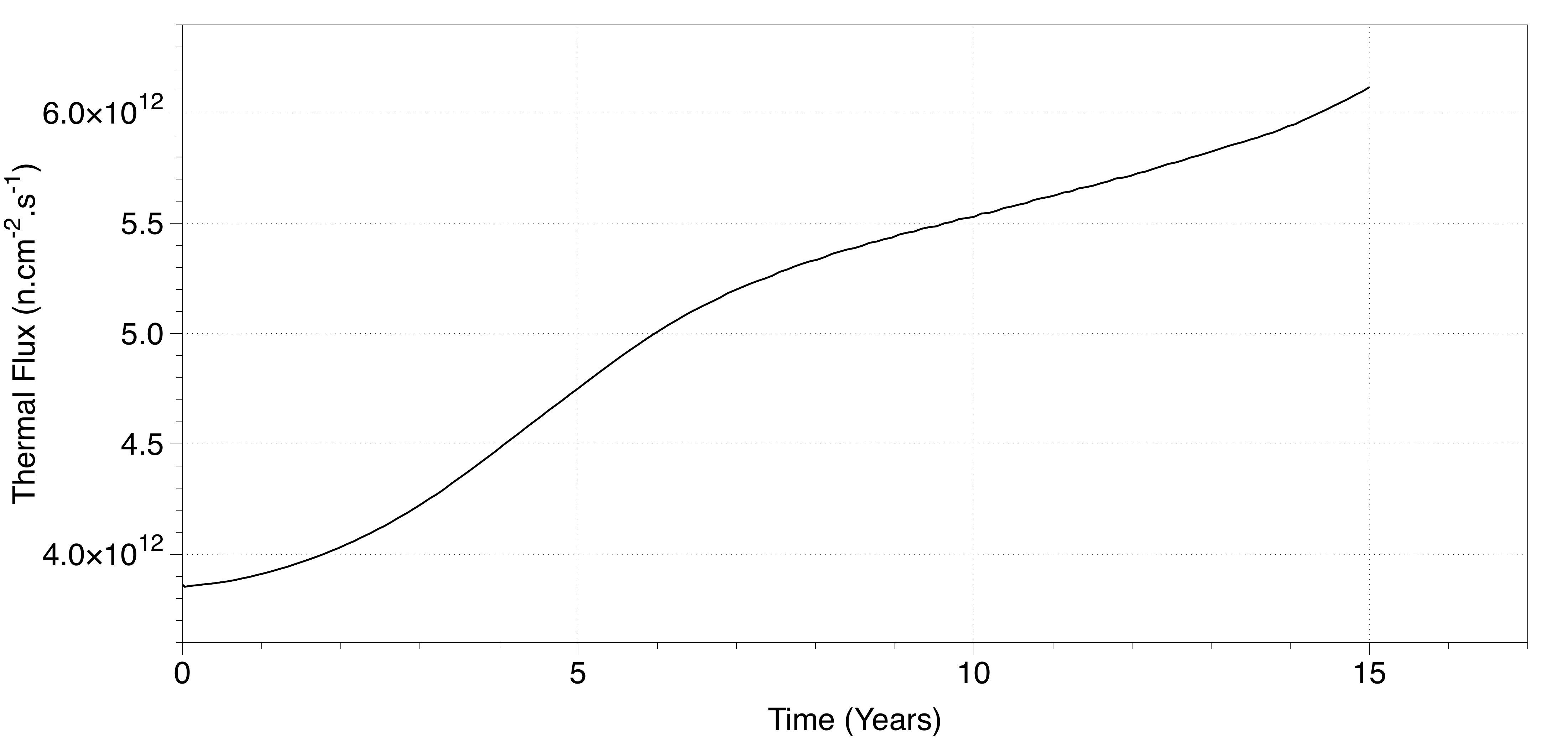}
\caption{Thermal flux as a function of time for the marine core designed in this study.}
\label{FLX_omega_core_8740}
\end{figure}

Figure \ref{FLX_omega_core_8740} shows the change in thermal flux over core life. Given the low power density of the core developed here (approximately 40\% of a standard PWR) and the much higher enrichment (around four times higher), it is expected that flux in the core will be much lower by around 0.4$\times0.25 = $ 0.1 times the flux of a conventional PWR. Note that the total thermal flux increases over core life and therefore depletion of the fuel - that is to say, to maintain a constant power output the flux must continuously increase. The larger thermal flux at the end of life (EOL) relative to beginning of life (BOL) helps explain the larger sensitivity to xenon at EOL shown in Table \ref{Xe_transients}.

To summarise, the necessity of a long core life, which results in low power densities, high enrichments and high poison concentrations also results in a low thermal flux throughout life and therefore limited $^{135}$Xe sensitivity. It should be noted that the low thermal flux for this core would carry a considerable economic penalty since $^{235}$U will less readily undergo fission, which results in the relatively high $^{235}$U enrichment and hence higher fuel costs. However, this needs to be weighed against the advantages associated with a longer core life that were highlighted earlier.

Besides overall core reactivity changes, the dependence of $^{135}$Xe on flux also creates the possibility for localised variations in $^{135}$Xe concentration within the core, which  can lead to so-called xenon oscillations. Xenon oscillations occur due to perturbations in the flux within a reactor (\cite{Duderstadt}).  Given the limited effect of xenon within the core (due to the low thermal flux) the marine core was found to be robust against axial and radial xenon oscillations.

\subsection{Monte Carlo Benchmarking}

Compared to conventional large PWRs the marine reactor in this study has a significantly lower power density (40~kW/l vs 105~kW/l) and significantly higher Gd$_2$O$_3$ (around 13 wt.\% vs 6 wt.\%) and $^{235}$U concentrations (around 13 wt.\% vs 4 wt.\%). CASMO-SIMULATE is well validated for conventional PWRs but the core designed in this study is expected to be considerably outside of the validation database CASMO-SIMULATE is based on. Therefore in this study comparisons between whole core Monte Carlo runs, using MONK-9A, and the nodal code SIMULATE-3 found discrepancies in the effective neutron multiplication factor throughout core life of between 1-2\%. Here discrepancy is defined as:

\begin{equation}
\rm{Discrepancy} = \frac{k_{\rm{MONK}} - k_{\rm{CASMO}}}{k_{\rm{MONK}}}
\label{disc_eqn}
\end{equation}
These discrepancies are much higher than those found when modelling conventional PWRs (discrepancies $<$~0.5\%) (\cite{Alexeev1998}) and should preferably be below 0.2\%.

In order to determine the likely reason for the discrepancies between SIMULATE-3 and MONK-9A, 2D lattice calculations (using the lattice geometry for the marine reactor) were performed in MONK and CASMO in order to benchmark the two codes; noting that any discrepancy at the lattice level (CASMO) will directly impact the accuracy at the nodal level (SIMULATE). Three cases were investigated; these were:

\begin{itemize}
\item Case 1 containing UO$_2$ fuel pins containing 5 wt.\% $^{235}$U surrounded by unborated light water;
\item Case 2 containing UO$_2$ fuel pins containing 13.5 wt.\% $^{235}$U surrounded by unborated light water; and
\item Case 3 containing UO$_2$ fuel pins containing 13.5 wt.\% $^{235}$U in addition to 76 burnable poison pins containing 12.75 wt.\% Gd$_2$O$_3$ surrounded by unborated water.
\end{itemize}

Case 1 represents a lattice equivalent to lattices found in conventional PWRs (albeit with the slightly larger diameter fuel rods of the marine core designed here - a conventional PWR will have a fuel rod diameter of around 0.95~cm \citep{WNI2012}, whereas the core developed here had a fuel rod diameter of approximately 0.98~cm, see Section \ref{sec:FuelPerform}). Cases 2 and 3 represent lattices with compositions that are typical of the marine core designed in this study.

In order to try and identify an underlying phenomenon that could be causing the discrepancy an attempt was made to quantify the hardness of the spectrum between the three cases. The parameter employed was k$_{\rm{fast}}$/k$_{\rm{total}}$, where k$_{\rm{total}}$ = k$_{\rm{fast}}$+k$_{\rm{thermal}}$ \footnote{In CASMO-4, k$_{\rm{fast}}$ and k$_{\rm{thermal}}$ are defined as the contributions to k$_\infty$ by neutrons with energies $\geq$ 4~eV and $<$ 4~eV respectively. k$_\infty=\frac{\Sigma_g\Sigma_i\upsilon\Sigma_{f,i}^g\phi_{i}^g V_i}{\Sigma_g\Sigma_i\Sigma_{a,i}^g\phi_{i}^g V_i}$ where $\upsilon$ is the number of neutrons released per fission, $\phi_{i}^g$ is the flux within mesh $i$ and group $g$, $V_i$ is the volume of mesh $i$, $\Sigma_{f,i}^g$ is the macroscopic fission cross-section of mesh $i$ and group $g$ and $\Sigma_{a,i}^g$ is the macroscopic absorption cross-section of mesh $i$ and group $g$.}.

{\footnotesize
\tabcolsep=0.11cm
\begin{table}[h!]
\centering
\begin{tabular}{|c|c|c|}
\hline
\bf{Case} & \bf{Discrepancy} & \bf{k$_{\rm{fast}}$/k$_{\rm{total}}$}\\ \hline
1 & 0.673 $\pm$ 0.006 & 0.254 \\ \hline
2 & 1.084 $\pm$ 0.011 & 0.435 \\ \hline
3 & 1.225 $\pm$ 0.017 & 0.543 \\
\hline
\end{tabular}
\caption{Discrepancies between MONK-9A and CASMO-4 for the various lattice types described in the text.}
\label{tab:lattice_discrepancies}
\end{table}%
}

Table \ref{tab:lattice_discrepancies} shows that the discrepancy significantly increases as the parameter k$_{\rm{fast}}$/k$_{\rm{total}}$ increases, implying that as the proportion of fission events occurs outside of the thermal region ($\geq$ 4~eV (\cite{Knott2010})) the discrepancy between these two codes increases.

In conventional PWRs around 80\% of fission events occur within the thermal region of the spectrum (\cite{Stammler}) and therefore light water deterministic codes tend to solve the neutron transport equation in more detail in the thermal region. This is the case in CASMO-4 where around 40 of the 70 energy groups are at thermal energies (\cite{CASMO_5}). Therefore, the number and distribution of energy groups is likely limiting. Furthermore, there is also the complication associated with the generation of group-averaged cross-sections at the beginning of the calculation, in particular those outside of the resonance region which often assume typical LWR flux distributions which are probably unsuitable for the reactor studied here.

To assess the dependence of discrepancy on group structure the code CASMO-5 was employed, which consists of 586 energy groups versus CASMO-4's 70 group structure (\cite{CASMO_5}). Whilst CASMO-5 has more energy groups in the fast region (around 170 groups, which in part also results in a higher computational demand than CASMO-4) as a thermal lattice code CASMO-5 still dedicates the bulk of energy groups to thermal energies. Discrepancies were found to be slightly smaller for Case 2 (0.8\% with CASMO-5 vs 1\% with CASMO-4) and Case 3  (1\% vs 1.2\%). Therefore, the group structure is important but these results indicate that the generation of group-averaged cross-sections at the beginning of the calculation (at the so-called library level) also play a significant role. To truly capture the dependency of cross-sections on neutron energy, around 10$^5$ energy groups would need to be used in a deterministic code (\cite{Knott2010}), which results in impractical demands on computational resource. This is why deterministic codes employ a coarser library structure, typically around $10^2$ energy groups, with the cross-sections for a particular interaction type (e.g. fission) at the library level calculated via:

\begin{equation}
\sigma_{\rm{g}} = \frac{\int_{E_1}^{E_2} \sigma (E) \phi (E) dE}{\int_{E_1}^{E_2} \phi (E) dE}
\label{group_xs}
\end{equation}
where $\sigma (E)$ is the cross-section from the continuous nuclear data library and $\phi (E)$ is a flux spectrum that is typical for group g, with group g spanning the energy range $E_1$ to $E_2$. Equation \ref{group_xs} requires an assumption to be made for the flux $\phi (E)$ within a particular group. The approximations employed in industry standard codes, such as CASMO, proves adequate for conventional LWR lattices (\cite{Stammler}) but as Table \ref{tab:lattice_discrepancies} indicates (the harder the spectrum the larger the discrepancy) likely has limited validity for the core designed here.

The inherent ability of Monte Carlo codes to more accurately represent the nuclear data and flux distribution throughout life results in a preference to regard MONK's results, using the continuous nuclear data library format, as less prone to systematic bias. As stated before, discrepancies of greater than 0.5\% between Monte Carlo and deterministic codes are generally considered large and would certainly bring into question the ability to predict the behaviour of the core throughout life to a sufficient standard to license the core design. However, this study is only preliminary and is focused on the general physical behaviour of the core over core life (such as xenon transient behaviour), where discrepancies of around 1 to 2 \% are not unduly burdensome. However, it highlights that deterministic codes will require modification to accurately predict the behaviour of the core studied here to the licensing standard employed for conventional PWRs.

Note that other uncertainties, including their corresponding effect on reactivity behaviour, will also be important. Design uncertainties, including temperature and geometry uncertainties, can be much larger than the observed 1-2\% discrepancy in \keff. It is important to note that in conventional PWRs the moderator temperature coefficient (MTC) can be most limiting, since at the beginning of life it can be only weakly negative when boron concentration is highest. Therefore, when uncertainties are included it is possible that MTC is no longer negative. However, in this core design - which operates without soluble boron - this will be less limiting since the MTC is more negative than in conventional PWRs.

\subsection{Fuel Performance}
\label{sec:FuelPerform}

A key element of core design is ensuring the survivability of the fuel during reactor operation. From this perspective a series of design criteria are normally prescribed, with the most limiting design criteria being:

\begin{itemize}
\item The extent of average clad hoop creep strain, which is typically limited to 1\%;
\item Ensuring the rod internal pressure does not increase to a level that could result in rupture of the cladding material. A conservative criterion typically applied is to limit the rod internal pressure to that of the coolant pressure (15.5 MPa); and
\item Limiting the extent of corrosion related phenomena (including oxidation and hydrogen embrittlement).
\end{itemize}

The rod internal pressure is usually limited to being below  the coolant pressure. This is because of the possibility that if the rod internal pressure is sufficiently high then the pellet-clad gap could re-open. Over time the initial pellet-clad gap will shrink in size due to a variety of phenomena including the fact that fuel pellets swell as they burn up. However, if the gap were to re-open later in a rod's life then this could quickly lead to clad rupture, due to the fact that the thermal conductivity across the re-opened gap will be heavily degraded by the presence of gaseous fission products Xe and Kr \citep{Uffelen2012}. Both Xe and Kr possess very low thermal conductivities and therefore a large temperature gradient across the re-opened gap can arise. This can result in excessive pellet temperatures, which accelerates the release of Xe and Kr from the fuel matrix, hence further degrading the thermal conductivity of the re-opened pellet-clad gap and increasing the temperature of the fuel, thus releasing more fission gas \citep{Uffelen2012}. Eventually the rod will rupture if its internal pressure becomes sufficiently high due to the large inventory of gaseous fission products. 

With respect to the clad hoop creep strain, a limit of 1\% is usually put in place to ensure that the likelihood of clad rupture is very low. As outlined in Section \ref{sec:operating_temp}, the low coolant temperature should limit corrosion-related phenomena; hence corrosion performance was not directly investigated here, as it is assumed that insisting on maximum clad external surface temperatures below $310^{\circ}$C will ensure the survivability of the fuel rod from a corrosion perspective.


The neutronic calculations carried out in the earlier section allow for the individual power histories for fuel rods to be determined. It was found that some fuel rods had EOL  burnups of approximately 120~GWd/tHM, with around 80\% of fuel rods having burnups below 100~GWd/tHM. Whilst oxide fuel exhibits many favourable properties (high melting point, low swelling rate and very good irradiation characteristics), and has been shown to operate to burnups $>$ 200~GWd/tHM in fast reactor environments, the existing database for validation of nuclear fuel under light water reactors is limited to around 100 GWd/tHM (\cite{Rossiter2011, Bremier2000}). Therefore, only the fuel performance of rods with burnups of less than 100 GWd/tHM are analysed here. It is assumed that future core iterations will reduce the range of burnups (perhaps via greater enrichment and poison zoning), thereby ensuring all rod burnups are within the existing validation database for LWR fuel rods and hence allowing greater confidence in accurately predicting rod behaviour.

The fuel rod parameters, which constituted the initial choice for an appropriate fuel rod design, are shown in Table \ref{Tab:Fuel_rod_params}.

{\footnotesize
\tabcolsep=0.11cm
\begin{table}[h!]
\centering
\begin{tabular}{|c|c|}
\hline
\bf{Parameter} & \bf{Value} \\ \hline
Pellet radius & 4.267$\times 10^{-1}$ cm \\ \hline
Pellet-clad gap & 8.200$\times 10^{-3}$ cm  \\ \hline
Clad thickness & 5.72$\times 10^{-2}$ cm \\ \hline
Upper plenum volume & 1.509$\times 10$ cm$^{3}$ \\ \hline
Lower plenum volume & 1.509$\times 10$ cm$^{3}$ \\ \hline
\end{tabular}
\caption{Fuel rod parameters.}
\label{Tab:Fuel_rod_params}
\end{table}%
}

Using the fuel performance code ENIGMA (v7.8) (\cite{Rossiter2011}) and sampling the fuel rods that achieved burnups up to 100~GWd/tHM it was found that peak clad hoop creep strains reached 4\%, significantly above the 1\% limit. However, modifications to the fuel rod design outlined in Table \ref{Tab:Fuel_rod_params} were investigated and found that increasing the pellet-clad gap size by a factor of 2.5 (from 8.2$\times 10^{-3}$~cm to 2.05$\times 10^{-2}$~cm) was sufficient to reduce peak clad hoop creep strain below the 1\% limit and due to the core's low power density, the greater gap size did not result in excessively high fuel centreline temperatures. It was found that for fuel rods up to 100~GWd/tHM, the initial chosen fuel design with plena lengths of 25.4~cm (10~inches) each at the top and bottom of the fuel, was sufficient to keep rod internal pressure below 15.5~MPa.

Furthermore, by using the ENIGMA fuel performance code, and sampling rods across the core, some fuel rods were found to have peak clad surface temperatures greater than the imposed limit of 310$^{\circ}$C. However, it was found that by varying the coolant mass flux across the core from around 1200~kg/m$^2$/s up to approximately 3600~kg/m$^2$/s was sufficient to limit peak clad temperatures to 310$^{\circ}$C. The disparities in peak clad surface temperatures across the core are a result of the larger peaking factors in the core due to elimination of soluble boron from the coolant. In the event further optimisation of the core design (such as through greater enrichment/poison zoning) is unable to reduce rod power history disparities, then a logical solution is to insist on variable channel flow across the core in order to minimise the differences in peak coolant temperatures. The use of variable channel flow is well-demonstrated in LWRs and is currently implemented both in BWRs and in some PWRs (of the VVER variety) through the use of zirconium-alloy boxes that surround the sides of the assembly (to stop cross-flow between assemblies) and through the incorporation of artificial blockages to vary channel flow rates between assemblies (\cite{Wiesenack2012}).

\subsection{Summary of final design values}

Table \ref{Tab:Core_params2} final design values based on the marine core developed here. The fuel rod parameters require further optimisation as outlined in Section \ref{sec:FuelPerform}.

{\footnotesize
\tabcolsep=0.11cm
\begin{table}[h!]
\centering
\begin{tabular}{|c|c|}
\hline
\bf{Parameter} & \bf{Value} \\ \hline
Number of fuel axial regions & 3 \\ \hline
Axial height of top and bottom fuel regions & 100~cm and 20~cm  \\ \hline
Number of RCCAs & 89 \\ \hline
RCCA type             & Hybrid hafnium-boron carbide \\ \hline
Pin diameter & 0.9842~cm \\ \hline
Upper plenum volume & 1.509$\times 10$ cm$^{3}$ \\ \hline
Lower plenum volume & 1.509$\times 10$ cm$^{3}$ \\ \hline
Inlet temperature         & 255$^{\circ}$C \\ \hline
Outlet temperature      & 285$^{\circ}$C \\ \hline
Peak clad surface temperature & 310$^{\circ}$C \\ \hline
\end{tabular}
\caption{Summary of core parameters for the marine core developed in this study.}
\label{Tab:Core_params2}
\end{table}%
}

\section{Conclusion}

This study has developed the core parameters for a marine reactor to be based on (including power output, capacity factor and coolant temperatures), with a particular emphasis on ensuring the parameters and chosen materials would realistically allow for the targeted 15 year core life to be achieved. Whilst the majority of goals have been achieved ($^{235}$U concentration below 20 wt.\% whilst achieving a 15 year core life, the elimination of soluble boron during operation and fitting a reactor core within a 3.5 m reactor pressure vessel), the current design requires further modification to ensure key fuel performance criteria are obeyed, in particular that the maximum clad hoop creep strain throughout life is limited to below 1\%. 

For the core design described here, relatively small modifications (in particular increasing the pellet-clad gap size) are required in order to ensure the rod internal pressure and maximum clad hoop creep strain meet their criteria. This is especially true if future core designs limit the maximum rod burnup to less than 100~GWd/tHM. The core has exhibited satisfactory shutdown margins, reactivity coefficients and favourable xenon transient behaviour, with the latter due to the low thermal flux in the core, which is due to the low power density and long core life. However, issues with current lattice codes to model hardened spectra LWR lattices have been identified. The advantages and disadvantages for soluble boron free PWR cores have also been investigated.

The core developed here represents the first workable design of a commercial marine reactor using conventional fuel, which makes realistic the idea of using nuclear reactors for shipping. The application of nuclear reactors for commercial shipping will have a number of non-technical barriers (including insurance, responsibility for waste management and the costs associated with maintaining a well-trained crew to operate the nuclear reactor), which we have outlined and where possible identified potential solutions to address these barriers.

\section{Acknowledgements}
Aiden Peakman's work was funded by an EPSRC grant (EP/G037140/1). The authors would also like to gratefully acknowledge input provided by Kevin Hesketh (National Nuclear Laboratory), Ian Palmer (retired) and Steve Walley (retired).

\section{References}

\bibliographystyle{unsrtnat}

\end{document}